# Advances in SPECT and PET Reconstruction for Theranostics: From Diagnosis to Therapy:

Kweku Enninful, BSc[a,1], Fardeen Ahmed, BSc[a,1], Bradley Girod, MD[b], Richard Laforest, PhD[b,c], Daniel L. J. Thorek, PhD[b,c], Vikas Prasad, MD[b,c], Abhinav K. Jha, PhD[a,b,c*]

[a]Department of Biomedical Engineering, Washington University, St. Louis, MO 63105, USA
[b]Mallinckrodt Institute of Radiology, Washington University, St. Louis, MO 63110, USA
[c]Siteman Cancer Center, Washington University, St. Louis, MO, USA
[1]These authors contributed equally
[*]Corresponding author: Department of Biomedical Engineering, Washington University, 1 Brookings Dr, St. Louis, MO 63105, USA. Email address: a.jha@wustl.edu

**Key points**:

- Growing interest in theranostic imaging applications has driven the need for and development of new reconstruction algorithms that address the unique challenges of these applications.
- We present advances in PET and single-photon emission computed tomography (SPECT) reconstruction for theranostics, highlighting multi-isotope reconstruction, prior-based reconstruction, projection-domain quantification and novel SPECT/PET instrumentation.
- We also highlight avenues for further research, including unmet clinical needs, imaging-system optimization, methods to extract maximal information from projections, and techniques for objective evaluation.

## Synopsis

The theranostic paradigm enables personalization of treatment by selecting patients with a diagnostic radiopharmaceutical and monitoring therapy using a matched therapeutic isotope. This strategy relies on accurate image reconstruction of both pretherapy and post-therapy images for patient selection and monitoring treatment. However, traditional reconstruction methods are hindered by challenges such as crosstalk in multi-isotope imaging and extremely low-count measurements data when imaging of alpha-emitting isotopes. Additionally, to fully realize the benefits of new imaging systems being developed for theranostic applications, advanced reconstruction techniques are needed. This review highlights recent progress and discusses critical challenges and unmet needs in theranostic image reconstruction.

## A. Introduction

Radiopharmaceutical therapies (RPTs), such as those based on Lutetium-177 (Lu-177), Radium-223 (Ra-223) and Actinium-225 (Ac-225), have attracted growing research interest as a cancer treatment modality to treat disseminated disease. In these therapies, a radioisotope is injected and delivered to the target cancer cells through proximity, or vectors that bind specifically to receptors on the cancer cells. Once delivered, the isotope emits alpha- (α-) or beta- (β-) particles, which can damage the DNA leading to cell cytotoxicity and effective local tumor control[1]. To ensure appropriate use of such targeted therapies, diagnostic imaging is often performed to confirm receptor expression and the presence of active binding sites prior to treatment. Post-therapy imaging also plays an important role in monitoring radiopharmaceutical uptake at disease sites and in radiosensitive organs. Quantitative assessment of this uptake is essential for estimating absorbed doses to tumors and normal tissues, which in turn enables treatment adaptation, prediction of therapeutic efficacy, monitoring of potential toxicities, and ultimately the personalization of therapy[2].

The theranostic paradigm provides a unique ability, in which a diagnostic radionuclide of a theranostic radionuclide pair, often positron emitting, is used to select patients and plan the RPTs based on the presence of the relevant receptors[2], while the therapeutic isotope of the pair, which attaches to the same receptors, is monitored for distribution during treatment. Some examples of these diagnostic and therapeutic pairs include Gallium-68 (Ga-68)/Ac-225, Ga-68/Lu-177, and Copper-64 (Cu-64)/Copper-67 (Cu-67). The diagnostic radionuclide can also be a gamma-emitter as is the case with Iodine-123 (I-123)/Iodine-131 (I-131), Iodine-124(I-124)/Iodine-131(I-131) and Lead-203 (Pb-203)/Lead-212 (Pb-212). The positron emitter can be imaged using positron emission tomography (PET). The therapeutic radionuclides typically emit gamma- or x-ray photons that can be imaged using single-photon emission computed tomography (SPECT). Thus, PET and SPECT provide a mechanism for visualizing and quantifying the isotope distribution. Crucially, however, these capabilities depend on robust image reconstruction methods.

**The need for developing new reconstruction methods for theranostic applications:**

Nuclear medicine imaging systems and reconstruction methods have been optimized for shorter lived isotopes with photopeak at 511 keV for PET (e.g. F-18 and Carbon-11(C-11)) or 140 keV for SPECT (Technitium-99m (Tc-99m)). With these isotopes, the typical image reconstruction procedure is to use the photopeak window data to estimate an image of the activity distribution over a three-dimensional voxelized grid. Even in this simple scenario, image reconstruction is confronted by challenges. These include photon non-collinearity and randoms in PET, low system sensitivity and collimator-introduced blurring in SPECT, attenuation and scatter of photons as they pass from the isotope to the detector, finite angular sampling during the image-acquisition procedure, the finite position and energy resolution of detectors, and Poisson noise statistics in both PET and SPECT. Compensation for these effects in image reconstruction in PET and SPECT is typically performed using ordered subsets expectation maximization (OSEM)-based algorithms, a variant of the maximum-likelihood expectation maximization (MLEM) algorithm. These algorithms account for Poisson photon-counting noise and allow modeling of the photon-propagation and imaging system physics[3]. The use of these algorithms has improved the accuracy of the reconstructed images, but development of improved methods continues to be an active area of research[4].

Image reconstruction in theranostics adds to the complexity. Often, therapeutic isotopes emit multiple imageable gammas as well as complex emission spectra including high energy photons and daughters. For example, Ac-225 decay produces the progeny Francium-221 (Fr-221) and Bismuth-213 (Bi-213) that also emits α-particles and gamma-ray photons. The gamma emissions from the daughters result in crosstalk contamination in the projections acquired from simultaneous multiple-isotope imaging due to factors such as the close proximity of gamma emission energies between isotopes, combined with the limited energy resolution of these imaging systems and the down scattering of photons from higher- to lower-energy windows. Crosstalk contamination degrades visual quality and quantitative accuracy[5], and thus methods that can handle these crosstalk contamination effects are needed.

The administered activities and imaging count rates in theranostics can also differ substantially from conventional diagnostic imaging. Particularly for α-particle RPTs, the administered activities are orders of magnitude lower than those of conventional diagnostic scans, leading to extremely low counts, which present substantial challenges for accurate assessment of radionuclide distribution. Also, at low counts, conventional voxel-based reconstruction is problematic, with even highly tuned protocols[6]. Further, at such low counts, even background contributions become significant and should be modeled.

To address these challenges, new systems are being developed for theranostic applications, incorporating innovations in system geometry, improvements in detector technology, and the development of hybrid imaging approaches. Examples include ring-shaped gantry systems with Cadmium Zinc Telluride (CZT) detectors[7,8] for SPECT, and [9]large-axial field-of-view (LAFOV) systems[9] for PET. As new systems are being developed, traditional reconstruction methods become limited in handling their complexities and capabilities.

Thus, new reconstruction methods tailored to these new technologies are needed to fully realize their clinical impact (Fig. 1).

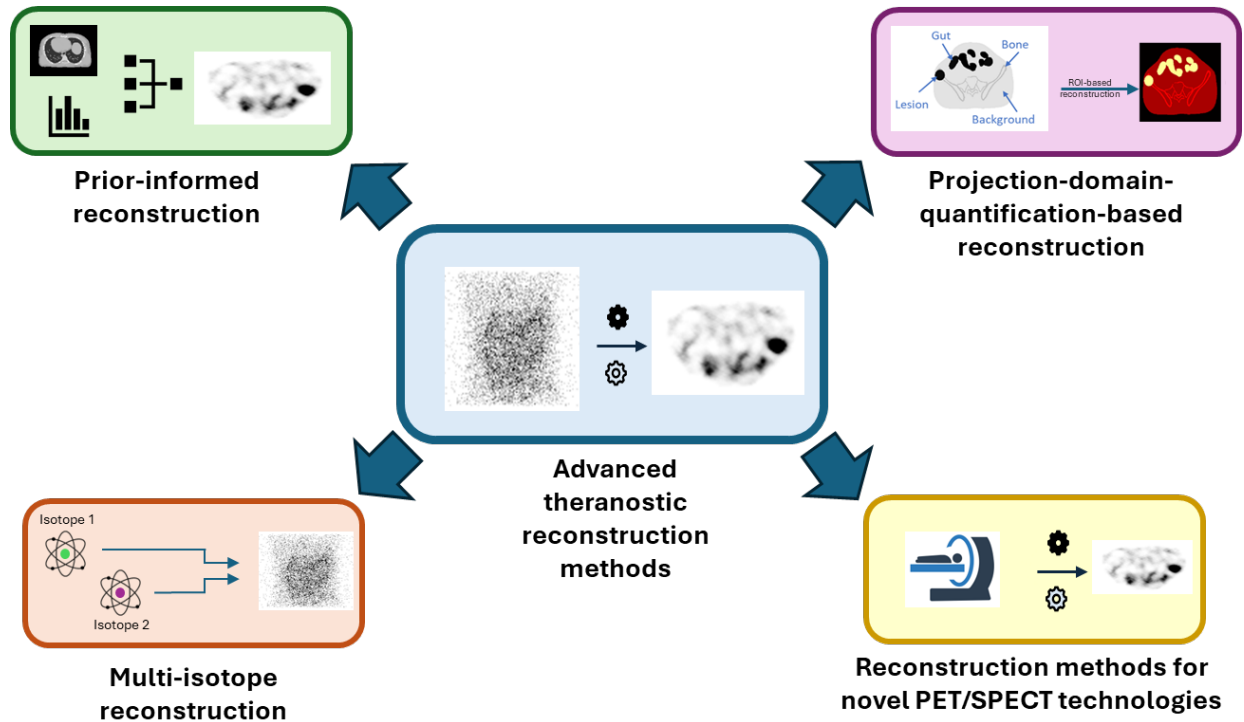

*Figure 1. Categories of advanced reconstruction methods discussed in this article.*

In summary, the evolving field of theranostic imaging presents unique challenges that conventional reconstruction methods are often ill-equipped to address. This has driven the development of new reconstruction algorithms. In this review, our goal is to survey these recent advancements in PET and SPECT reconstruction methods for theranostic applications and discuss important challenges and unmet clinical needs in this field.

**B. Advancements in computational SPECT and PET for theranostics**

**B.1 Methods for multi-isotope reconstructions and quantification**

Radioisotopes used in theranostic applications often have a complicated gamma-emission spectra where the parent isotope may decay into one or more generations of radioactive daughter isotopes. Each of these decays can produce charged particles, that affect dose, and photons. The photons thus produced have a variety of energies and can produce cross-talk in the various imaging energy windows. Depending on the relative half-lives and the location of the radionuclide at the time of decay, these daughters can have independent biodistributions that may be important in accurately assessing absorbed dose or modeling the crosstalk during image reconstruction[10]. Thus, accurate estimation of the activity distribution of the parent and daughter isotopes requires methods for multi-isotope reconstruction. Moreover, such multi-isotope reconstruction methods enable assessing multiple molecular targets within a single imaging session.

SPECT systems allow for acquisition of projections in multiple energy windows. This provides the data needed for joint reconstruction of two or more radioisotopes. However, the limited energy resolution of SPECT systems and processes such as scatter in the patient and in the SPECT system result in crosstalk contamination between the projection data collected in different energy windows[5]. To address these challenges, several approaches have been proposed. One such method is the dual isotope OSEM-based (DOSEM) reconstruction[11], which uses a model-based approach for estimating and correcting crosstalk (Fig. 2), including downscatter and Pb X-ray contamination[12]. The the effective source scatter estimate (ESSE) technique has been used to model downscatter[13] and collimator Pb X-ray model was used to estimate Pb X-ray contamination[14]. Alternatively, Monte Carlo (MC) simulation has been used to model and estimate down-scatter contributions, yielding more accurate modeling of crosstalk[15,16]. This approach has been validated in simultaneous In-111/Tc-99m [15] and Ho-166/Tc-99m imaging[16]. MC-based methods are, however, computationally expensive and require substantial computational resources to perform within clinically acceptable computation times[17], making it inaccessible to low-resource centers. Other methods for crosstalk correction include independent component analysis[18], artificial neural networks[19,20], penalized maximum likelihood algorithms[21], and the analytical photon distribution-based iterative reconstruction method[22]. These methods have been validated for Iodine-123 (I-123)/Tc-99m imaging[19,20,22], motivating use for other theranostic applications. Another advancement is the development of dual-isotope projection-domain quantification methods, which have been used to jointly estimate regional activity uptake for Th-227/Ra-223 [23]. Recently, simultaneous estimation of more than two isotopes using a generalized multi-isotope estimation method has shown efficacy in reliably estimating the regional activity uptake of Ac-225, Fr-221and Bi-213[24]. These methods, which will be described later, have shown that performing joint quantification yields more reliable estimates compared to estimating regional activity for each isotope individually.

In PET, even if multiple positron-emitting isotopes are present, the positron-electron interaction yields photons at the same 511 keV energy, limiting the ability to image multiple isotopes. However, recent advancements have provided avenues for multi-isotope imaging by using high-energy prompt-gamma photons that are emitted immediately after positron decay and are isotope specific[25]. By collecting the prompt gamma photons alongside the conventional 511KeV annihilation photons, we can perform spatially accurate image reconstruction that is independent of the positron range. Andereyev and colleagues have proposed the dual-isotope PET (DIPET) approach, where one of the radioisotopes is a pure positron emitter and the other emits an additional prompt gamma[25]. They developed a dual isotope expectation maximization (EM) reconstruction algorithm that demonstrated the ability to simultaneously estimate the distribution of the two PET radionuclides[26] and have since expanded that method to develop a generalized EM algorithm, which includes data from acquisitions obtained from staggered injections of the radiotracers[27]. Results indicate that combining both acquisitions obtained from staggered injections and acquisitions using positron-gamma emitters improves performance visually and quantitatively[27]. Recently, another generalized reconstruction method, multiplexed PET (mPET), has been proposed[28]. This method can separate two PET isotopes within a single PET acquisition based on the type of gamma emission alone and has shown promise in simultaneous imaging of I-124 and Zr-89, and I-124 and Ga-68 in mice [28]. The use of a PET-SPECT-CT scanner, Versatile Emission Computed Tomography (VECTor)/CT, equipped with a high-energy cluster-pinhole collimator has been proposed for preclinical imaging[29]. Here, the reconstruction method models and corrects for crosstalk between isotopes by scaling and subtracting contributions based on MC simulation [29]. A new paradigm uses quantum entanglement filtering to reduce random coincidences between prompt and annihilation photons, maximizing the benefits of prompt gamma radiotracers and improving reconstruction accuracy [30].

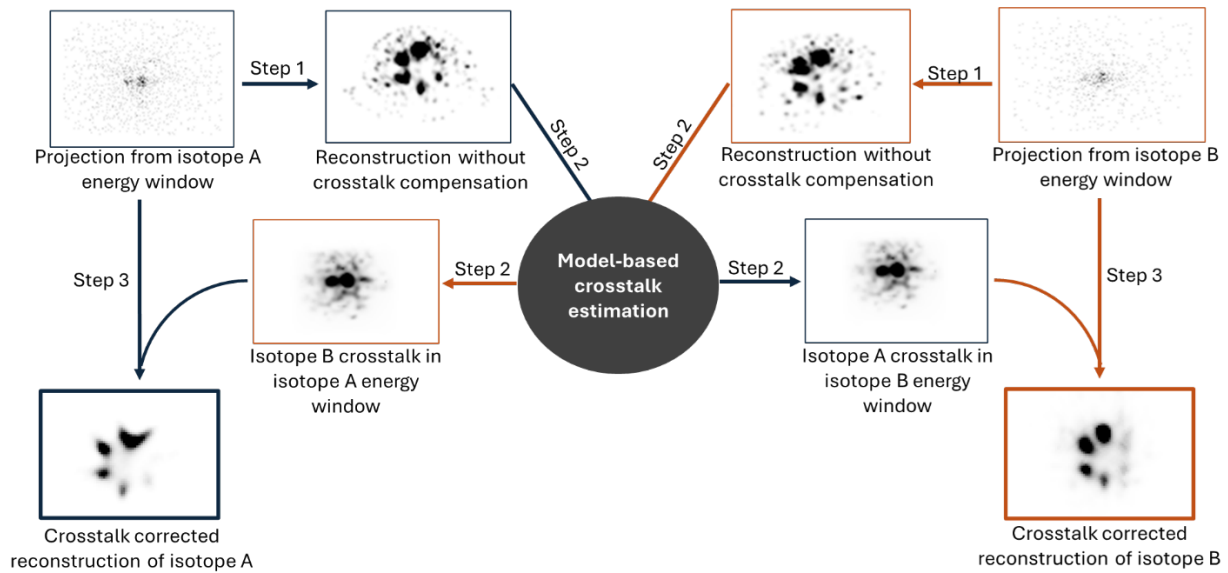

*Figure 2. Flow chart of the dual-isotope OSEM-based reconstruction method using the model-based technique, to estimate crosstalk.*

**B.2 Prior-informed reconstruction methods**

The literature on prior-informed reconstruction is extensive, and in keeping with the scope of this paper, we focus only on those methods that have been proposed for theranostic applications. In the estimation-theory literature, a prior refers to an assumed distribution that represents beliefs about the estimated parameters[31]. In image reconstruction, a prior serves the purpose of imposing spatiotemporal constraints on the estimated images, guiding the reconstruction process towards solutions that are more likely to be realistic and physically plausible[32].

Priors can typically be divided into handcrafted and learned priors. Handcrafted priors encode analytically specified assumptions about the properties of the isotope distribution. The effect of priors is adjusted through fine-tuning of additional parameters, such as the weight and neighborhood size of the prior. A common hand-crafted prior is the smoothing prior, which has been used in studies to reconstruct Ga-68 PET scans, achieving improved contrast and noise properties[33,34]. However, a smoothing prior may yield over-smoothed images[35].

Hand-crafted priors can also include information from other modalities, such as CT images, making the priors patient specific[36–38]. The availability of dual-modality systems such as PET-MRI, PET-CT and SPECT-CT has enabled near-real-time incorporation of anatomical priors in reconstruction. Dewaraja and colleagues demonstrated that the incorporation of boundary information from CT-based anatomical prior can yield images with reduced bias and noise, and improved preservation of boundaries between organs[37]. Other studies have explored the use of functional priors for image reconstruction in theranostics, particularly, pre-therapy diagnostic PET scans as priors. Marquis and colleagues demonstrated the use of Ga-68 and Cu-64 scans for post-therapy reconstruction of Lu-177 and Cu-67 SPECT images, respectively, capitalizing on the similarity in distributions of the diagnostic and therapeutic radiopharmaceuticals[39,40]. Simultaneous incorporation of both functional and anatomical priors has also been investigated[41] (Fig. 3).

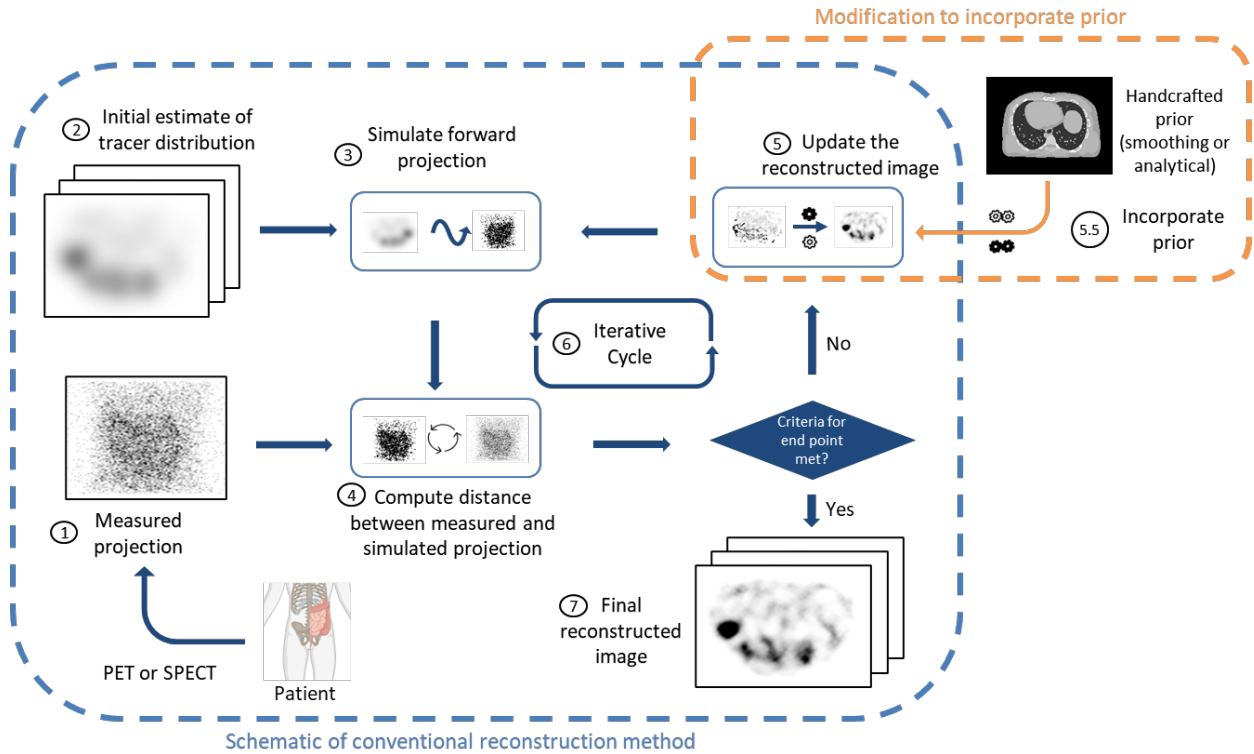

*Figure 3. Schematic of an example method for incorporating handcrafted priors into conventional iterative reconstruction methods.*

Learned priors are image-population-derived priors learned from imaging datasets, typically using deep-learning (DL)-based methods. These methods have gained strong interest driven by advances in computational resources and data availability[42]. Most methods in the theranostics space can broadly be categorized into the following categories (Fig. 4):

- **Attenuation and scatter-compensation methods**: DL-based attenuation compensation methods can be broadly categorized into direct methods - training models to predict attenuation-compensated images from non-attenuation compensated images[43,44], and indirect methods - training models to generate the attenuation map for reconstruction[45]. Similarly, DL-based scatter compensation methods address scatter either directly during the reconstruction process[46], or indirectly by estimating scatter projections before reconstruction[47,48]. DL-based scatter compensation has been of interest in whole-body PET, due to the increased scatter and random events from the larger axial length of the field-of-view[49,50]. When estimating the scatter projections, existing studies have either used MC simulations [47] or attenuation map-based sinograms to derive the scatter data [48].

- **Post-reconstruction image refinement**: In these methods, a DL model is trained to predict higher-count/high-resolution images given lower-count/lower-resolution

- reconstructed images. This has been investigated for noise reduction[51,52] and partial volume correction[53]. Deng and colleagues demonstrated the use of DL-based methods to reduce Ga-68 dose in PET[54].
- **Projection-domain processing**: In these methods, DL approaches are used to process or supplement projection data prior to reconstruction. For example, in Lu-177 SPECT, synthetic intermediate projections generated by DL were added to the measured projection data[55–57], which was then processed by OSEM for reconstruction. However, OSEM-based methods assume positivity constraint and Poisson-distributed noise, which may not hold with DL-processed projections[58]. Further, the theoretical derivation of the OSEM procedure assumes a forward model that encompasses all operations performed on the projection data. Addressing these limitations is a crucial area for future research.

Studies have also explored other types of DL-based methods. Liu and colleagues developed a conditional generative adversarial network (GAN)-based method that directly estimates the reconstructed image from the measured projection data[59]. Experiments with simulation data showed that the model outperformed conventional reconstruction methods in terms of bias, variance and noise robustness. However, when applied to a smaller clinical dataset, the method reportedly failed to depict anatomical details clearly, compared to the results from filtered backprojection. These observations show that while end-to-end reconstruction is promising, they face notable limitations, particularly the need for larger and more diverse clinical datasets. Further, given the well-established understanding of imaging system physics, some argue that relying on DL to replace these physical models entirely may not be prudent. Additionally, studies have looked at combining handcrafted priors with DL. For example, CT-derived priors have been incorporated through DL-trained regularizers to reconstruct Lu-177 DOTATATE images from shortened scans[60]. These methods have shown improvements in noise reduction and quantification when compared with conventional regularized methods.

While the use of DL for reconstruction has demonstrated promise, further research is needed. Incorporating learned priors in reconstruction can be challenging given requirement of large, diverse, clinically representative patient datasets and significant computational resources[61]. DL-based methods also face the risk of hallucinations and artifact generation, resulting in inaccurate representations in the reconstructed images[62] and limiting performance on clinical tasks[63]. Further, DL models based on visual fidelity metrics, such as pixel-wise mean squared error (MSE), may not yield improved performance on clinical tasks[63]. An additional complexity is the difficulty in detecting and interpreting these defects[64]. However, given the potential of DL to learn prior information from patient populations, developing approaches to address these challenges is a promising area of research. As an example, studies have shown that by training DL algorithms that model specific clinical tasks such as defect detection, performance on those tasks can be improved[65]. Thus, designing DL algorithms that specifically model clinically relevant theranostic tasks is an exciting area of research.

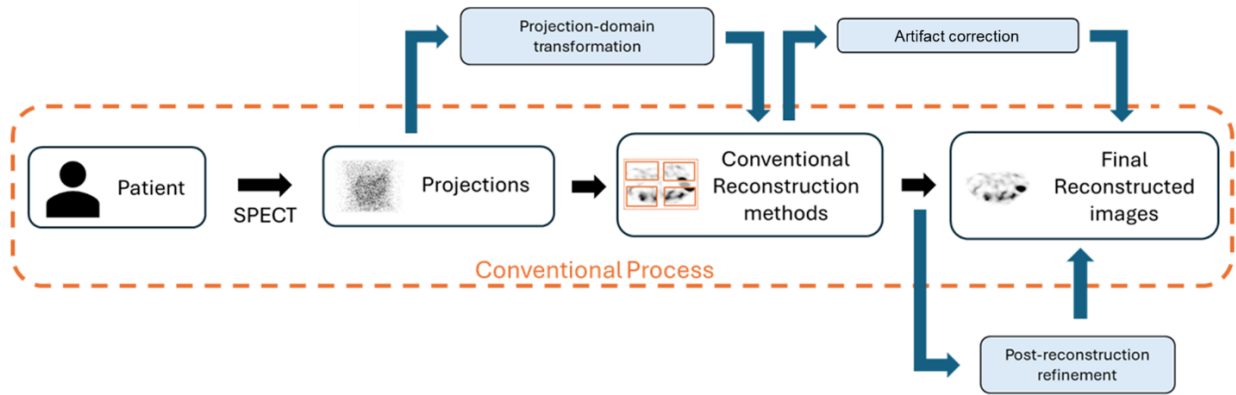

*Figure 4. Schematic illustrating the various DL-based reconstruction methods to incorporate learned priors*

**B.3 Projection-domain quantification (PDQ) methods**

In theranostic imaging applications involving β-particle therapies or PET, the number of detected counts in projection data is higher than that in imaging of α-RPTs using SPECT. In the latter, the number of detected counts in projection data can be extremely small due to the low administered activities combined with the limited SPECT system sensitivity[66,67]. In these cases, voxel-based reconstruction approaches are challenged by the need to estimate many voxel values from highly noisy measurements. The methods are observed to have high biases [66,68,69] (19-35%) and imprecision[67] in estimating activity even with highly fine-tuned protocols. One approach to addressing this issue is to reduce the number of parameters that need to be estimated. In this context, for organ and lesion-based dosimetry applications, the clinical task from SPECT images is estimating the mean uptake in those volumes of interest (VOIs). Thus, methods that can estimate the regional isotope uptake directly from the projection data could provide a mechanism to address the low-count issue.

In early work, Carson and colleagues proposed a method that directly quantifies the activity uptake within VOIs from projection data, thus skipping the reconstruction step[70]. Inspiration from that idea has led to several methods proposed for α-particle SPECT, where the VOIs can be defined, for example, from the low-dose CT that is acquired in conjunction with SPECT. A low-count quantitative SPECT (LC-QSPECT) method[67] was designed to estimate mean uptake in specified VOIs for a single isotope using projection data obtained from a single energy window. In this method, the imaging physics and spectra of the emission isotope were modeled using (MC)-based processes[67]. The background contributions in the projection data, termed as stray-radiation-related noise, which becomes relevant at the low count levels, was also modeled. LC-QSPECT has been validated for Ra-223-based α-RPTs in single-scanner studies in both realistic simulations and with physical phantoms[67], demonstrating high accuracy and precision. The method demonstrated good reproducibility across multiple scanner-collimator configurations, as evaluated in a virtual imaging trial (VIT) referred to as in silico imaging trial for quantitation accuracy (ISIT-QA)[71].

LC-QSPECT has been extended to a multiple-energy-window projection-domain quantitative SPECT (MEW-PDQ) method that can reliably estimate the regional uptake of two isotopes, taking advantage of projection data from multiple energy windows to improve performance[72]. MEW-PDQ has been validated for the joint quantification of Th-227 and Ra-223, exhibiting high accuracy and precision in estimating the mean activities of both isotopes. Studies have also shown that the method exhibits reliable performance when there are moderate levels of misalignment between CT and SPECT scans[72]. Further, this method also been recently generalized to a multi-isotope low-count quantitative SPECT (MI-LC-QSPECT) method that estimates the regional uptake of any number of γ-emitting isotopes present in specified VOIs[24]. MI-LCQSPECT has been applied in Ac-225-based α-RPT SPECT, providing reliable quantification for Ac-225 and its daughters, Fr-221 and Bi-213 [24]. Crosstalk among the isotopes is modeled in both MEW-PDQ and MI-LC-QSPECT methods. Other proposed methods include EQ-Planar [73], which uses defined VOIs and whole-body planar projections to estimate organ activities and dynamic time activity curves [74]. Another approach has been proposed to estimate the activity and uncertainty in segmented VOIs by combining a Monte-Carlo-based likelihood fit and projection data obtained from a highly sensitive collimatorless detector[75]. This has been validated for Ac-225 in spherical and mouse phantoms and simulations[75].

The PDQ methods assume the distribution of activity within a VOI is homogenous, which may not be the case[76]. Studies have shown that while these methods can tolerate small amounts of heterogeneity, larger spatial heterogeneity deteriorates performance. Towards addressing this issue, a strategy was recently proposed to account for intra-regional uptake heterogeneity, namely, the Wiener-estimator-based PDQ (WIN-PDQ) method[77]. This method requires prior knowledge of the intra-regional uptake heterogeneity model and the distribution of parameters that determine heterogeneity. The method was evaluated on MC-simulations of anthropomorphic phantoms with Ra-223 uptake and achieved nearly unbiased activity uptake estimates for an ensemble. We note here, though, that WIN-PDQ was developed in the context of image-quality evaluation studies, and not necessarily for reconstruction[77]. Nevertheless, this provides an avenue for further research. Another research area is integrating PDQ and voxel-based reconstruction-based methods, providing a mechanism to include complementary advantages of visualization of intra-region activity distribution and accurate estimation of activity. For example, a voxel-based SPECT reconstruction can be used in conjunction with the CT scan to provide definitions of the different VOIs for the PDQ method. This could provide an even more reliable approach to clinically implement this method, using a framework as presented in Fig. 5.

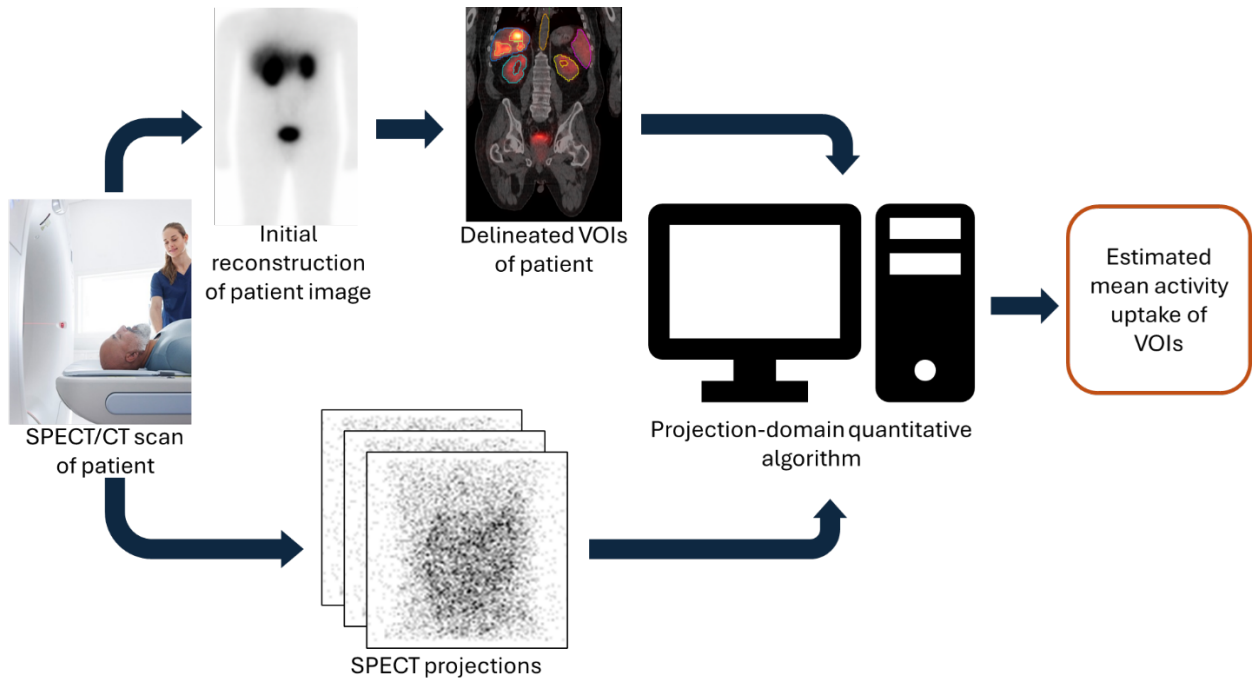

*Figure 5. A potential framework for clinical implementation of projection-domain quantitative SPECT approaches that use the SPECT and CT reconstructions to define the VOIs.*

**B.4 Reconstruction methods for novel PET/SPECT technologies**

New adaptive geometry SPECT systems with Cadmium Zinc Telluride (CZT) detectors placed in a ring geometry, such as the GE StarGuide[7] and the Spectrum Dynamics Veriton[8], have been investigated for theranostic applications. In a whole-body post-therapy SPECT/CT with (Lu-177)-DOTATATE and (Lu-177)-PSMA-617 study, Song and colleagues demonstrated that fast SPECT acquisitions (total scan time of 12 minutes) are feasible with the GE StarGuide system, as compared to GE Discovery 670 Pro SPECT/CT (total scan time of 32 min)[78]. Given the adaptive geometries of this system and the use of CZT detectors, advances in reconstruction are desirable to use the potential of CZT technology fully. A recent study showed that a block sequential regularized expectation maximization (BSREM) algorithm outperformed the manufacturer implementation of OSEM when imaging Lu-177 with StarGuide[79]. The BSREM algorithm is known to reach convergence without the detriment of noise-induced image degradation, resulting in less noisy images[80]. However, since these detectors are CZT-based, tailing-related crosstalk can result in loss of energy resolution, and methods to compensate for those are much needed [81]. Also, the effect and modeling of septal penetration with these systems requires further investigation.

Other novel SPECT instrumentation designs for theranostic applications have also been proposed that could benefit from reconstruction advances. This includes a recently proposed sensing collimator imager (SCI)-SPECT system [82] that incorporates the geometry of both a conventional pinhole collimator and coded aperture; replacing the heavy metal collimator with scintillator crystals to increase sensitivity and improve resolution[82]. Additionally, a collimator-less SPECT system has been introduced for whole-body imaging of Ac-225[83], and Alpha-SPECT-Mini, a small-pixel Cadmium Telluride (CdTe) detector-based imaging system capable of imaging α-emitters such as Ac-225 has also been proposed[84]. Developing tailored reconstruction algorithms for these systems will be crucial in exploiting their capabilities for theranostics.

A major advance in PET system instrumentation has been the advent of large-axial field-of view (LAFOV) systems such as the total-body Explorer[9], PennPET-Explorer[85], and Biograph Vision Quadra PET/CT[86]. While these systems open new research and clinical frontiers, they require larger computational resources and are subject to increased scatter and random events[50]. Towards addressing the issue of computational complexity, studies have developed graphics processing unit (GPU)-accelerated reconstruction methods[87,88], which can be applied in theranostics applications for further research. Another research area is developing DL-based reconstruction methods for LAFOV PET systems[43,50]. A trained DL model that accounts for the increased scatter in LAFOV PET systems can speed up the reconstruction process and have been demonstrated to improve scatter correction to yield better reconstruction performance[50]. To effectively evaluate these DL methods, it is necessary to compare their performance with other specialized reconstruction methods such as BSREM, and TOF-PET OSEM. Lindstrom and colleagues compared BSREM and TOF-PET[89] in the context of LAFOV-PET, but to-date, no studies have compared these methods with DL-based LAFOV reconstruction methods. Further development and appropriate evaluation of these DL-based LAFOV PET reconstruction methods is an important area of research.

**C. Discussion**

Image reconstruction for theranostic applications is an emerging field of high clinical relevance offering stimulating research opportunities. In this review, we have outlined four domains where notable progress has been made, also outlining further research areas in these domains. Nonetheless, given its relative nascency, the field presents numerous other avenues for future investigation. We discuss some of these avenues here (Fig. 6).

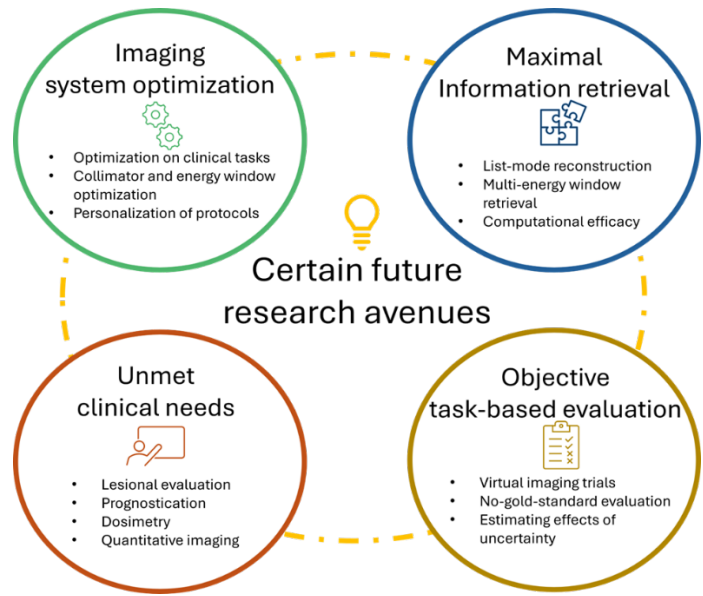

*Figure 6. A schematic illustrating certain avenues for further research in the area of image reconstruction for theranostic applications.*

## C.1 Strategies for imaging-system optimization

Reconstruction methods are limited by the quality and information of the measured data; thus, strategies that can facilitate collecting the most informative photons for the clinical task are needed. Towards this goal, methods have been proposed for task-specific optimization of the collimator design[90], energy window width[91], and a combination of collimator hole width and energy window width[92]. Other parameters to optimize could include bed position[93], acquisition time[94] and time spent per projection[95].

Performing such optimization requires developing techniques to extract task-specific information from the measured data. For organ/lesion dosimetry, the clinical task is estimation of activity in radiosensitive organs and lesions. For this purpose, the WIN-PDQ estimator described above, which is optimal in that it minimizes the ensemble mean square error between the estimated and true regional uptake values[77], enables designing acquisition protocols for improved performance on this task[96]. Similarly, ideal observers can be used to guide system design when the objective is to maximize the information content in projection data for detection tasks. Instead, if the goal is to enhance detection performance on reconstructed images as assessed by human readers, then human-observer studies may be needed. However, these studies are time-consuming, expensive, and logistically challenging. At earlier stages, anthropomorphic model observer-based studies may provide a mechanism to identify promising optimized imaging system configurations for human-observer studies[97]. Finally, the clinical task is often a combination of detection and quantification. Thus, another essential research frontier is methods to optimize imaging systems for joint detection and quantification tasks.

While most research on task-specific imaging system optimization has focused on optimizing the system for populations, another emerging paradigm is to adapt the imaging-system design for specific patients. Barrett and colleagues outlined this idea in the context of adaptive design of SPECT systems[98,99]. Studies have shown that such personalization can lead to more accurate performance on the task of defect detection[100]. Expanding on these efforts in the theranostics field to optimize the design of imaging systems and protocols is another exciting research frontier.

**C.2 Reconstruction methods for maximal information retrieval**

As outlined earlier, a central challenge in the SPECT of α-emitters is the extremely low number of detected counts. Reliable quantification in these conditions can be significantly facilitated if the reconstruction method efficiently uses most of the detected photons and extracts the maximum possible information from each detected photon. One approach is to process data from multiple energy windows. For α-particle SPECT, this can substantially increase the detected counts, effectively improving system sensitivity [67]. Multi-window acquisitions can also theoretically yield more precise estimates of regional activity uptake, as observed in SPECT studies with Y-90[101] and Th-227[23]. Further, using singular value decomposition-based approaches to model scatter in multiple energy windows can improve noise characteristics in scatter-corrected reconstructions[102,103].

Another approach is to develop methods for reconstruction with list-mode (LM) data. Processing data in LM format provides a mechanism to extract the maximal information content from each photon. In the LM data format, the attributes of each detected photon such as the position of the interaction in the scintillation detector, the energy deposited by the detected photon in the detector, and the time of detection can be recorded (Fig. 6). Typically, these attributes are binned, which leads to information loss. Processing data in LM format has been shown to improve reconstruction performance[104,105], and performance in estimation tasks[106]. Studies have also shown that the energy attribute may contain information to perform attenuation compensation[107]. In fact, Guérin and colleagues observed in PET-based simulation studies that incorporating the energy attribute while performing scatter compensation reduced the bias in the activity distribution estimates by up to 40% for a single-scatter simulation model[108].

Combining the above two approaches, Rahman and colleagues developed a LM multi-energy window low-count SPECT reconstruction method for isotopes with multiple emission peaks[109]. Simulation-based experiments for estimating Ra-223 activity uptake demonstrated that the proposed method achieved improved performance in estimating regional activity compared with approaches using a single energy window or binned data. The method was developed for 3D SPECT and could model multiple orders of scatter but was evaluated in 2D and with single-scatter settings, primarily due to the high computational requirements. The results motivate continued research in developing and validating methods that can process LM data over multiple windows, and in investigating strategies to make this approach feasible for standard clinical workflows.

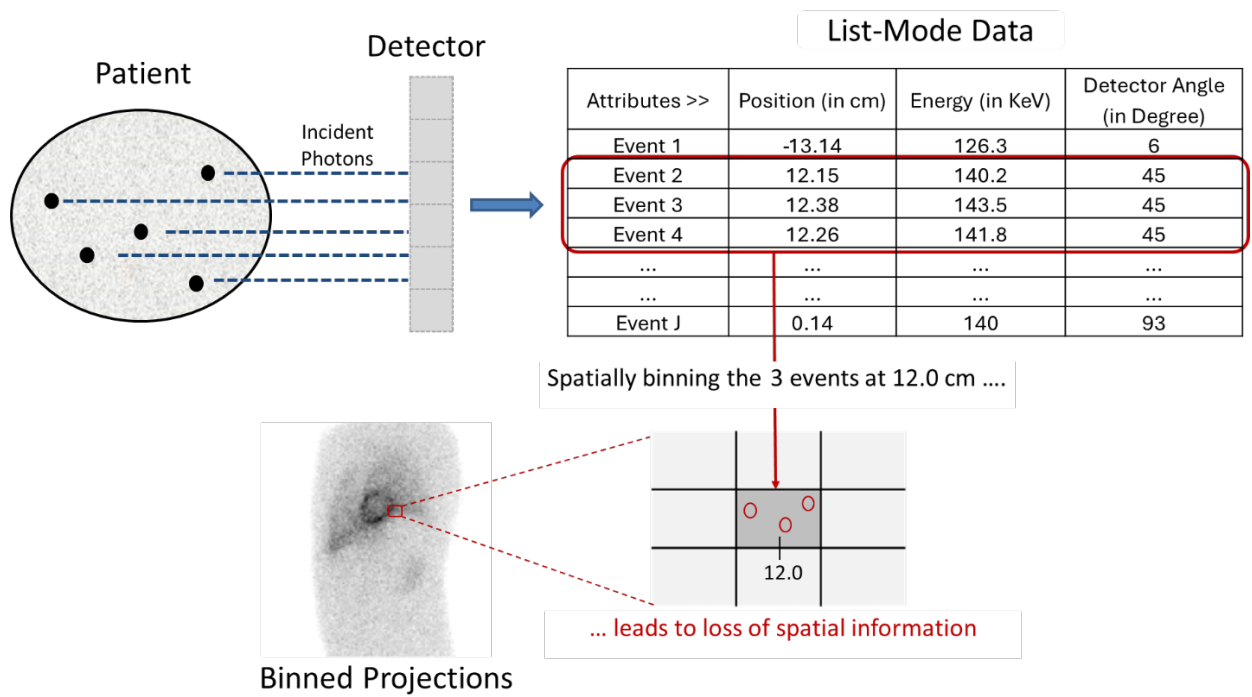

*Figure 7. A schematic illustrating the loss of information in transitioning from LM to binned data*

**C.3 Techniques for objective task-based assessment of reconstruction methods**

With the advances in development of reconstruction methods for theranostics, an important research area is developing objective, clinically relevant strategies for assessment of these methods. Since the reconstructed images are acquired for specific clinical tasks, eventually this assessment should be based on performance in those tasks. Thus, developing clinical task-specific validation methodologies is an area of important investigation.

Liu and colleagues highlights four emerging frameworks for objective evaluation of quantitative imaging methods[110] that are applicable to assessing theranostic reconstruction methods for quantitative tasks such as dosimetry. One key development we highlight is VITs[111]. These provide a low cost, relatively inexpensive, and controlled mechanism to evaluate reconstruction methods with known ground truth and identify the most promising methods for further clinical validation. For example, Li and colleagues used VITs to demonstrate the efficacy of their quantitative SPECT methods for Ra-223[71] and demonstrated the reproducibility of that method across nine different scanner-collimator configurations. Further, they used VITs to show the reliability of another method for joint quantification of Th-227 and Ra-223 [23]. Additionally, VITs may provide a means to optimize costly and laborious multi-institutional trials (the gold standard for modern evidence-based medicine) to make them more efficient and more likely to succeed.

For clinical translation of reconstruction methods, validation with clinical data is important. However, such validation requires ground truth, which is typically unavailable. Towards addressing this issue in the context of quantitative tasks, no-gold-standard evaluation (NGSE) techniques have been developed[112,113]. These techniques assume a linear relationship between the true and the measured quantitative values, characterized by linear-relationship parameters, normally distributed noise, and that the true quantitative values have been sampled from some unknown but parametric distribution. Using maximum-likelihood methods, these techniques estimate these linear relationship parameters without access to ground truth. An NGSE technique was developed to assess quantitative SPECT reconstruction methods for I-131 imaging. The technique demonstrated the ability to yield an accurate ranking of these reconstruction methods on the task of estimating regional uptake in VOIs without the availability of ground truth [112]. There is ongoing research on further advancement and validation of this technique[114,115] and this is an important area of future research.

Developing clinical task-based evaluation strategies is even more essential in the era of AI since AI-based methods are known to generate images that may visually look appealing but not improve performance on the underlying clinical task[63]. To this effect, the SNMMI AI Task Force has proposed the RELAINCE guidelines for evaluation of AI algorithms for nuclear medicine, recommending that evaluation of AI algorithms yield a claim that quantitatively describes the performance of the algorithm[116]. Also, a framework for task-based evaluation of AI-based medical imaging methods has been outlined[117] which advocates for physicians to have a key role in performing such evaluation studies.

Another need is for methods to estimate the impact of photon-noise introduced uncertainty on performance in clinical tasks. Polson et. al have proposed a computationally efficient algorithm that propagates uncertainty from projection data through clinical reconstruction algorithms, quantifying uncertainties in total activity within VOIs[118]. The method has been validated on Lu-177 and Ac-225 phantoms and patient data and has been made publicly available in the PyTomography reconstruction library[119] and 3D Slicer[120].

**C.4 Unmet clinical needs**

Numerous challenges remain in image reconstruction in theranostics. Lesional evaluation for the assessment of disease heterogeneity is improving with full-body PET/CT scanners but will require a focused effort to improve assessment for different clonal populations and changes during therapy. Multi-isotope imaging has and will continue to play a role in disease heterogeneity and will be improved upon by new tracers and potentially multiplex PET/CT, so new methods in this direction will be needed. Similarly, methods for optimizing SPECT/CT dosimetry protocols, including the number and timing of post-treatment assessments, will be necessary to ensure reliable dosimetry while maintaining the ability to yield accurate results.

Most importantly, quantitative imaging is poised to enable better selection of patients for combination treatment, and reliable quantification will continue to require advances in reconstruction. Cancer lesions that are less likely to respond to RPT because of sub-optimal radiation absorbed dose may benefit from combination with other targeted therapies such as checkpoint inhibitors and olaparib, that either augment the RPT mechanism of action or provide additional anti-cancer effects, specifically to inherently radioresistant cells. Another interesting concept is to combine radiopharmaceuticals with different radionuclide properties e.g. combination of β-particle emitting radionuclides with α-particle radionuclides. RPTs also have great potential in modulating the tumor microenvironment, thereby converting an immunotherapy non-responsive lesion into an immunotherapy-sensitive one. Because of the high objective response rate shown in several prior RPTs, several ongoing trials are exploring the possibility of bringing the RPT into the first line of therapy. As toxicities of RPT manifest themselves sometimes years after the treatment is over, it is imperative to have quantitative tools to reliably predict the outcome (toxicity and response) so that patients are not devoid of a second or subsequent line of treatment because of, e.g., lack of adequate bone marrow or renal function. Again, advances in reconstruction would be needed to obtain reliable quantification.

**D. Conclusion**

Given the rising prominence of RPTs, and the vital role of image reconstruction in both pre- and post-therapy settings, there is an urgent need for advanced reconstruction methods tailored for theranostic applications. In this paper, we present an overview of current reconstruction methods developed and evaluated for theranostic applications and discuss areas of future research. Overall, we envision this article as a guide to understanding the current landscape of advanced reconstruction methods and identifying the unmet needs in this field. By further advancing the identified research areas, the various clinical processes in RPTs can be enhanced, resulting in improved therapeutic outcomes in cancer patients.

**Acknowledgements**: Financial support for this work was provided by the National Institute of Biomedical Engineering and Bioimaging R01 EB031962, R01 EB031962-03S1 and the National Science Foundation CAREER award (Grant Number: 2239707).

**Disclosures**: The authors have nothing to disclose.